# Towards a Comprehensive Bibliography for SETI

Alan Reyes[1], Jason Wright[1]

*areyes@psu.edu, jtw13@psu.edu*

[1]Department of Astronomy & Astrophysics and Center for Exoplanets and Habitable Worlds, The Pennsylvania State University, 525 Davey Laboratory, University Park, PA 16802, USA

## Abstract

In this work, we motivate, describe, and announce a living bibliography for academic papers and other works published in the Search for Extraterrestrial Intelligence (SETI). The bibliography makes use of bibliographic groups (bibgroups) in the NASA Astrophysics Data System (ADS), allowing it to be accessed and searched by any interested party, and is composed only of works which have a presence on the ADS. We establish criteria that describe the scope of our bibliography, which we define as any academic work which broadly: 1) advances knowledge within SETI, 2) deals with topics that are fundamentally related to or about SETI, or 3) is useful for the better understanding of SETI, and which has a presence on ADS. We discuss the future work needed to continue the development of the bibliography. The bibliography can be found here or by using the bibgroup field (bibgroup: SETI) in the ADS search engine.

## 1 Introduction

The Search for Extraterrestrial Intelligence (SETI) has been a formal pursuit of generations of scientists stretching back to the mid-20th century. Although thousands of works have been published on the subject, it has been difficult to manipulate or peruse the corpus of published SETI work due to the absence of a public bibliography. The goal of this work is to deliver the first step towards such a bibliography, by starting with those papers already cataloged on the ADS [1].

There are several reasons why a SETI bibliography is warranted and would be an important asset for the field.

The first is that it has been somewhat difficult to trace the field's historical development or natural progression of ideas, especially in the case of those ideas that were either not published in an academic journal or even at all. One example of this is the confusion around the proper citation for the Drake Equation - often given an improper citation or no citation at all. The earliest edited and citable reference to a work by Drake containing the equation (as identified by Drake himself in private communication) is an excerpt from the 1965 textbook published by the Oxford University Press, ``The Radio Search for Intelligent Extraterrestrial Life''. For a long time, there was no ADS entry for this reference, and hence no citation available from the ADS BibTeX generator. (Upon communication with the ADS administration, an entry has since been created [2].)

There is also a lack of uniformity in the jargon used by SETI practitioners. Having a definitive bibliography would help authors to trace the emergence of specific terms and understand their context and precise meanings. This in turn would foster the consolidation of present and future use of SETI jargon.

More broadly, some other arguments in favor of a bibliography include:

1. SETI has historically had a difficult struggle with public perception, especially at the legislative level. When politicians - with the ability to set budgets and allocate funding for scientific endeavors - have a gross misunderstanding of what SETI is, and even more so of the level of academic rigor adhered to by its practitioners, the field suffers. This perception has resulted in the wholesale defunding of SETI-related research. Having a comprehensive bibliography would help to formalize the field and improve public perception, and reintroduce the possibility of funding.

2. For a field as small as SETI, it ought not be difficult to keep tabs on all publications in an organized manner. The low publication rate and small overall literature volume makes a comprehensive bibliography a realistic goal.

3. A bibliography would enable projects to quantify the amount of searched parameter spaces (e.g., in the context of carrier waves) [3, 4] by providing a complete record of past observations.

4. Concrete criteria for what constitutes a "SETI paper" helps users understand which works are and are not expected to be cataloged.

5. A bibliography would enable the generation of BibTeX citations for the SETI community. Since attribution is a fundamental aspect of the scientific process, having an easy way to identify relevant citations and extract their bibliographic records promotes engagement with the literature and facilitates the connection and progression of ideas.

Prior to this work, no well-maintained compilation of SETI works universally accessible by any interested party and with the explicit goal of completeness has existed. Several partial compilations have been made, but none were easily available to the public. In many instances, access was held privately and only available as, for instance, a PDF file upon request from the maintainer. The lack of uniform formatting between compilations is also an impediment to their usability. In spite of these shortcomings, they served as a healthy place for us to begin the development of the ADS bibgroup.

In §2, we discuss the specific goals of this work, with emphasis on its scale and intended outcomes. In §2.1, we set the vetting criteria for papers or other citations to be considered viable additions to the bibliography. Then in §3, we discuss the exact procedure that was undertaken to produce the bibliography. Lastly, in §4, we reflect on what has been accomplished and what work still needs to be done.

**2 Goals**

Since a bibliography which is both cross-disciplinary and comprehensive across all media was too ambitious for our efforts, we restricted our scale to focus on only those academic works which are directly relevant to the astronomical practice of SETI, according to a general criterion established below. The end product is a SETI bibliographic group (bibgroup) hosted on the NASA ADS. Bibgroups are search filters that can restrict search results to query only those records attached to a particular ADS collection. They are also a powerful way to discover thematic groupings in databases and generate impact metrics, among other functions. We choose to host the bibliography on ADS because of the close relationship between SETI and traditional astronomy, which means that many SETI works already have entries in the ADS. Additionally,  ADS automatically ingests much of the arXiv, so that any publications there—even those SETI papers which are not directly connected to astronomy—can be easily added.

While in the process of collating citations, we encountered and took note of relevant citations in formats other than traditional academic and peer-reviewed publications. Alongside these were citations to SETI works outside of the physical sciences (e.g., the social ramifications of a first contact, etc.) While a comprehensive bibliography would include these entries, for practical reasons we have restricted ourselves to those with a presence on the ADS and leave their inclusion for a future project.

An important auxiliary goal of this work is to develop a list from which an informal SETI "canon" can be derived. The intention is that this list covers a broad range of topics in and around SETI, which would give the interested reader a healthy introduction to the field. This resource could then be used as the basis for the curriculum of a graduate course on SETI, such as [the new graduate course at Penn State](#).

2.1 Criteria

An important question to tackle when attempting to construct a comprehensive bibliography is: ``What constitutes a SETI paper or work?'' There must be a rigorous selection criteria which allows for the decisive categorization of works. Towards this end, Table 1 lists some broad areas or classes of papers that were identified while processing the bibliography, separated into what is considered in-scope and out-of-scope by this bibliography.

Most obviously, those papers which discuss formal observational campaigns, apparati, and results are included. Importantly, null results are included, as they often demonstrate the level of scientific rigor involved in SETI, as well as contribute to the overall searched parameter space. Papers which attempt to formalize or sharpen the Fermi Paradox, as well as those which offer solutions to it (many of which go by ``[insert] hypothesis/scenario''), are also included. Artifact SETI is a branch of SETI which takes the approach of searching for evidence of the activity of extraterrestrial intelligent civilizations as opposed to their transmissions [5]. As it is its own sub-field, publications from artifact SETI necessarily merit inclusion. Also to be included are those citations which deal with SETI on a "meta" level, namely those papers or other media about SETI as a field rather than on a topic within SETI. Those citations that made it to our collection of foundational papers but which were not already circumscribed by the other categories are added by default due to their pedagogical relevance (e.g. [6]). Based on these categories, the criterion for a citation to be included in the bibliography can be summarized as:

"Any work which either: 1) advances knowledge within SETI, 2) deals with topics that are

Table 1: Defining the Scope of the Bibliography (see text for details)

| *In-scope:* | *Out-of-scope:* |
|---|---|
| SETI Observation papers and null results | "Pure" Astrobiology |
| SETI-specific instrumentation papers | Physics or General Radio Instrumentation |
| Fermi Paradox and attempted solutions | Crank/Pseudoscientific Hypotheses |
| Artifact SETI theory and searches | Sociology/Anthropology/Humanities Papers (for the purposes of the current effort) |
| Meta-SETI | |
| Papers that made it to our collection of foundational papers | |

fundamentally related to or about SETI, or 3) is useful for the better understanding of SETI."

Excluded types of citations include papers on what is dubbed "pure" astrobiology - although SETI is fundamentally a part of astrobiology, the formal study of life in the universe, it is mostly concerned with the search for evidence of technological life, and hence technosignatures. (Generally speaking, SETI is concerned with the search for any form of intelligent life, but a non-technological intelligent species may not leave detectable traces.) Therefore, by this distinction, "pure" astrobiology encompasses all other works on life in the universe, on such topics as the abiogenesis, panspermia, or the search for alien biosignatures, to name a few examples, and is excluded from a "pure" SETI bibliography.

Another class of paper frequently cited by SETI papers are papers from physics journals. SETI searches are made possible by contributions from fundamental physics. However, while they are certainly relevant for understanding the context of any given SETI approach, a general physics paper is not necessarily appropriate for inclusion to the bibliography. Similarly, papers describing general radio facilities or instruments which do not primarily serve SETI are likewise excluded, even if they have SETI time allotments on them.

As with many fields, SETI is not immune to an inflow of crank or quack papers. Indeed, due to the level of popular interest on the topic, SETI is especially susceptible to them. While it is important to be inclusive of a variety of unconventional ideas, some carefully considered limitations ought to be set on how deviant a paper can be from the traditional scientific approach. We identified a handful of such citations that were deemed inappropriate for inclusion in what is supposed to be a collection of scientific SETI papers. Moving forward, careful vigilance will be observed to guarantee the scientific authenticity and integrity of included works.

Lastly, as previously discussed, we have not incorporated papers outside the physical sciences. Some excluded papers were those on post-biological evolution, animal communication, and speculative linguistics or reply construction. While these are important and relevant works, they have poor representation on the ADS and are outside the scope of this primary effort. Finding ways to incorporate them, whether here or in some other repository, would be an interesting future extension of this project.

We also encountered a variety of media types, including but not limited to papers, popular articles, books, book reviews, poster abstracts, talk or presentation abstracts, conference proceedings, private letters or correspondences, videos and video books, lecture slides, and government documents. Many of the aforementioned media formats do not constitute academic works and thus are outside the scope of this compilation. Therefore, a second layer of criteria needs to be established to distinguish between academic and non-academic media, which can often times have blurred boundaries, as in the case of non-peer-reviewed works. For the purposes of this bibliography, only papers, books, and abstracts (which also obeyed the first criterion) were admitted without further consideration. While it may be beneficial to have repositories of SETI in other media, it may be difficult to track down every instance of a SETI-related work in those media, and therefore should be considered another ambitious task for a future work.

## 3 Procedure

Our foundation was an ADS library we constructed from a citation tree of SETI papers, which originally had ~1900 entries in it, many of which did not fit the inclusion criterion. The point of this cursory assemblage was to put together an unrefined library of all potential SETI citation candidates, and only then later refine its constituents via a robust vetting procedure. To build it up further, we conducted a manual search of citations based on SETI-themed keywords appearing in title or body, allowing for a further several hundred candidates to be added. An example of one kind of search we performed was for citations which contained any combination of the body keywords ``extraterrestrial intelligence civilization'' using the OR logic of the ADS search system. Another search we performed was for titular keywords ``Fermi paradox'' using AND logic. (The aforementioned are just two examples of a variety of combinations of commonly used SETI words.) We purposefully cast a wide net so as to be as inclusive as possible, and to catch those missed by the citation tree.

Next, we folded in partial compilations generously provided by Robert Gray, Shelley Wright, Jill Tarter, and others by searching for entries in their lists on the ADS. The largest of these compilations was a multi-hundred page file maintained by Stephane Dumas which contained records on papers, popular books, proceedings, reports, IAC conferences, special journal issues, a popular magazine, and miscellaneous meetings. Another large compilation was a one hundred page NASA JPL reference publication consisting of a mixed variety of papers, popular articles, and government documents. Lastly, all entries maintained by the SETI.news website were added. We were unable to complete a search for every entry on all of these lists, but believe that we have found the vast majority with a presence on ADS.

The next task was to vet and condense the library. The first step in this process was to export the ADS citations to a spreadsheet where they could be assigned a variety of disposition flags. We allocated a total of eight distinct disposition categories. Going individually through the list of candidates and vetting them based on their title, several hundred citations were deemed suspect and flagged for a further manual inspection of their abstract. On the second pass, we perused the abstracts and eliminated entries whose content were deemed outside our scope, as well as many entries with missing abstracts, duplicate titles, and/or broken links.

Lastly, we created a new SETI bibliography of ADS entries matching our criteria, and requested that ADS create a bibgroup based on it. A static image of this library (as opposed to the dynamic and living bibgroup) can be found [here](#).

## 4 Discussion and Future Work

In general, we pursue completeness as an aspirational goal. The effort strives to be as high fidelity a representation of the body of academic works published in SETI as is possible given the tools and resources available as well as the accepted selection criteria. However, in the case of more obscure or the newest works, the bibliography will not have perfect coverage, and it can be expected that some percentage slip through and are yet to have a presence. That is why the bibliography will be continually updated as more citations are discovered, and we welcome suggestions from the community for new entries in the form of a link to the appropriate ADS entry sent via email to [astrowright@gmail.com](mailto:astrowright@gmail.com). We will hold the editorial privileges for the library and properly vet all suggested additions to maintain the integrity of the master list.

Other future work for this project includes completing the folding-in of the previous compilations, the continued tab-keeping on SETI works from other non-ADS-covered disciplines as well as media formats, and the continued integration of [SETI.news](#) [7] publications as they are released. For popular articles, an online spreadsheet shall be a starting point for documenting publications of that format. Importantly, we intend to collaborate in the future with librarians to create bibliographic entries using professional software for non-ADS entries. Similarly, the creation of ADS entries for relevant papers lacking an ADS presence is warranted and is currently underway by the ADS administrators.

We hope that this bibliography will be a useful resource for current and future researchers in the field that fosters the utilization of the rich SETI literature.

A popular summary of this work is [here](#).The bibgroup is directly linked [here](#).

## Acknowledgements

This paper grew out of a [final project](#) in the 2018 graduate course on SETI at Penn State. The authors are extremely grateful to Stephane Dumas, Jill Tarter, and Robert Gray, whose partial compilations set the foundation of this work. We also thank Frank Drake, Alberto Accomazzi, Carolyn Grant, Shelley Wright, Andrew Siemion, Robert Gray, James Guillochon, and Michael Oman-Reagan for their personal correspondences, advice, and information which aided this work.

## References

[1.] M.J. Kurtz, G. Eichhorn, A. Accomazzi, C.S. Grant, S.S. Murray, and J.M. Watson, ``[The NASA Astrophysics Data System: Overview](#)'', *Astronomy and Astrophysics Supplement*, **143**, 41-59, 2000.

[2.] F.D. Drake, ``The Radio Search for Intelligent Extraterrestrial Life'', *Oxford University Press*, 323-345, 1965.

[3.] J.C. Tarter et al., ``SETI turns 50: five decades of progress in the search for extraterrestrial intelligence'', *Proceedings of the SPIE*, **7819**, id:781902, 2010.

[4.] J.T. Wright, S. Kanodia, and E. Lubar, ``How Much SETI Has Been Done? Finding Needles in the n-dimensional Cosmic Haystack'', *The Astronomical Journal*, **156**, 260, 2018.

[5.] J.T. Wright, S. Sheikh, I. Almár, K. Denning, S. Dick, and J. Tarter, ``Recommendations from the Ad Hoc Committee on SETI Nomenclature'', *arXiv e-prints*, arXiv:1809.06857, 2018.

[6.] J. M. Cordes et al., ``Theory of Parabolic Arcs in Interstellar Scintillation Spectra'', The *Astrophysical Journal*, **637**, 346-365, 2006.

[7.] J. Davenport and D. LaCourse, http://SETI.news, (Last Accessed 10th December 2018)